\documentclass[final,3p,times]{elsarticle}

\usepackage{amsmath,amsthm,txfonts,graphicx}

\newcommand{\pa}{\partial}

\renewcommand{\Re}{\operatorname{Re}}
\renewcommand{\Im}{\operatorname{Im}}

\newcommand{\RR}{\mathbb{R}}

\newcommand{\al}{\alpha}
\newcommand{\be}{\beta}
\newcommand{\ga}{\gamma}
\newcommand{\de}{\delta}
\newcommand{\ep}{\varepsilon}

\newcommand{\la}{\lambda}

\renewcommand{\th}{\theta}
\newcommand{\ph}{\varphi}

\newcommand{\ola}{\overline{\lambda}}

\newcommand{\opsi}{\overline{\psi}}

\newcommand{\bU}{\mathbf{U}}
\newcommand{\bV}{\mathbf{V}}

\theoremstyle{plain}

\theoremstyle{definition}

\usepackage{color}
\newcommand{\Dmitry}[1]{{\color{blue}{\texttt Dmitry: #1}}}

\begin{document}

\begin{frontmatter}

\title{Lump chains in the KP-I equation}

\author[labela]{Charles Lester}
\author[labelb,labelc]{Andrey Gelash}
\author[labeld]{Dmitry Zakharov}
\author[labelb,labele]{Vladimir Zakharov}
\address[labela]{Department of Physics, University of Arizona, Tucson, AZ}

\address[labelb]{Skolkovo Institute of Science and Technology, Moscow, Russia}
\address[labelc]{Institute of Automation and Electrometry SB RAS, Novosibirsk, Russia}
\address[labeld]{Department of Mathematics, Central Michigan University, Mount Pleasant, MI}
\address[labele]{Department of Mathematics, University of Arizona, Tucson, AZ}


\begin{abstract}
We construct a broad class of solutions of the KP-I equation by using a reduced version of the Grammian form of the $\tau$-function. The basic solution is a linear periodic chain of lumps propagating with distinct group and wave velocities. More generally, our solutions are evolving linear arrangements of lump chains, and can be viewed as the KP-I analogues of the family of line-soliton solutions of KP-II. However, the linear arrangements that we construct for KP-I are more general, and allow degenerate configurations such as parallel or superimposed lump chains. We also construct solutions describing interactions between lump chains and individual lumps, and discuss the relationship between the solutions obtained using the reduced and regular Grammian forms.



\end{abstract}


\end{frontmatter}



    

\section{Introduction}

The Kadomtsev--Petviashvili equation 
\begin{equation}
[u_t+6uu_x+u_{xxx}]_x=-3\alpha^2u_{yy}
\label{eq:KPI}
\end{equation}
was derived in~\cite{1970KadomtsevPetviashvili}, and was first mentioned by its current name in~\cite{1974ZakharovShabat}. The KP equation is the subject of hundreds of research papers and several monographs~\cite{1991AblowitzClarkson,2018Kodama,1993Konopelchenko,1980NovikovManakovPitaevskyZakharov,2010Osborne}. The KP-I and KP-II forms of the equation are physically distinct and correspond to $\alpha^2=-1$ and $\alpha^2=1$, respectively. 

The KP-I and KP-II equations are universal models describing weakly nonlinear waves in media with dispersion of velocity. However, from a mathematical point of view they are quite distinct. They have numerous physical applications, such as the theory of shallow water waves (see, for instance, the monographs~\cite{1991AblowitzClarkson,2018Kodama}) and plasma physics (Kadomtsev and Petviashvili were both renowned plasma physicists). Both KP-I and KP-II are Hamiltonian systems. The Cauchy problem for both equations is uniquely solvable  for initial data in $L^1$ (see~\cite{1983FokasAblowitz,1981Manakov,1990Zhou}). However, KP-II is completely integrable, while KP-I, in general, is not (see the paper~\cite{1991ZakharovSchulman} for the analysis of the difference between the two equations).

Both versions of the KP equation are solvable using the inverse scattering method. The KP equation is the compatibility condition for an overdetermined linear system
\begin{equation}
\alpha\Psi_y+\Psi_{xx}+u\Psi=0,\quad
\Psi_t+4\Psi_{xxx}+6u\Psi_x+(3u_x+3\alpha w)\Psi=0,\quad w_x+u_y=0.
\label{eq:Lax}
\end{equation}
The Lax representation for KP was found independently by Zakharov and Shabat~\cite{1974ZakharovShabat} and Dryuma~\cite{1974Dryuma}. For KP-I we have $\alpha=i$ and Equation~\eqref{eq:Lax} is a non-stationary one-dimensional Schr\"odinger equation with the potential $-u$, while $\alpha=1$ corresponds to KP-II, and the linear problem is a heat equation with a source term. This fact alone reveals the substantial underlying difference between the theories of KP-I and KP-II.

The KP-I equation has a rich family of rational solutions, describing the interactions of stable, spatially localized solitons known as \emph{lumps}. A lump solution of KP-I was first constructed numerically by Petviashvili~\cite{1976Petviashvili}, who developed an original method for numerically constructing stationary solutions for a wide class of nonlinear PDEs. Lumps and their interactions were first studied analytically in~\cite{1977ManakovZakharovBordagItsMatveev}, and received their name in~\cite{1979SatsumaAblowitz}, where they were constructed using the Hirota transform. Krichever~\cite{1978Krichever,1979Krichever} showed that the dynamics of the lumps in KP-I is controlled by the Calogero--Moser system. Lumps with distinct asymptotic velocities retain their velocities and phases after scattering, but lumps with the same velocity undergo anomalous scattering, and may form bound states known as multilumps~\cite{1993GorshkovPelinovskyStepanyants,2018HuHuangLuStepanyants,1994Pelinovsky,1998Pelinovsky,1993PelinovskyStepanyantsA,2017Stepanyants}. Lump and multilump solutions of KP-I were described in the framework of the inverse scattering method in~\cite{2000AblowitzChakravartyTrubatchVillarroel,1999VillarroelAblowitz}.

Unlike the KP-I equation, KP-II is not known to have spatially localized solutions, nor does it have nonsingular rational solutions. Instead, the KP-II equation has an interesting family of \emph{line-soliton} solutions. An individual line-soliton is a translation-invariant traveling wave. When several line-solitons interact, they form complicated evolving polyhedral arrangements \cite{1978AnkerFreeman,2007Biondini,2006BiondiniChakravarty,2003BiondiniKodama,2008ChakravartyKodama} that are described by an elaborate combinatorial theory (see~\cite{2014KodamaWilliams} and the monograph~\cite{2018Kodama}). Line-soliton solutions also exist for KP-I but are unstable with respect to transverse perturbations; this was shown in the original paper~\cite{1970KadomtsevPetviashvili} for large perturbations and in~\cite{1993PelinovskyStepanyantsB,1975Zakharov} for all scales. For stability of three-dimensional solitons, see~\cite{1982KuznetsovTuritsyn}.

The goal of this paper is to describe a new class of solutions of the KP-I equation, which we call \emph{lump chains}. A simple chain consists of lumps evenly spaced along a line, with the lumps propagating with a single velocity at an arbitrary angle to the line. Lump chains can interact by splitting or merging, and the large-scale structure of lump chain solutions of KP-I resembles that of the line-soliton solutions of KP-II. However, lump chains may have degenerate behavior that does not occur with line-solitons, such as parallel and superimposed chains. More generally, we construct solutions consisting of lump chains that emit individual lumps, which may be then absorbed by other chains. Solutions of KP-I containing a periodic chain of lumps have been described by a number of authors~\cite{1985Burtsev,2010GorshkovOstrovskyStepanyants,1993Konopelchenko,1993PelinovskyStepanyantsB,1983Zaitsev,1984ZhdanovTrubnikov}. However, to the best of our knowledge, solutions consisting of several interacting lump chains have never been considered before. 

We construct solutions of KP-I using the Grammian form of the $\tau$-function. This form was derived using the dressing method in~\cite{1974ZakharovShabat}, and using Sato theory in~\cite{1989Nakamura}, and is perhaps less known than the Wronskian form. The dressing method was first used to solve the KdV equation in the pioneering paper~\cite{1973Shabat}, and was generalized and applied to the KP-II equation in~\cite{1974ZakharovShabat}. A more modern treatment can be found in the papers~\cite{1990Zakharov,1985ZakharovManakov,1979ZakharovShabat}.

As we have noted, individual lump solutions of KP-I are stable, while line-solitons and lump chains are unstable. In~\cite{1993PelinovskyStepanyantsB} it was shown that a line-soliton can emit a lump chain, hence the latter should be considered as an intermediate stage of the instability development. In the long run, a line-soliton transforms into an expanding cloud of lumps, which can be treated as a model of integrable turbulence.

\section{The Grammian form of the $\tau$-function}

\label{sec:Grammian}

The purpose of this paper is to study a family of solutions of the KP-I equation that can be constructed using the Grammian form of the $\tau$-function, which we now recall~\cite{1989Nakamura,1980NovikovManakovPitaevskyZakharov,1993PelinovskyStepanyantsB}. Fix a positive integer $M$, which we call the \emph{rank} of the solution. Let $\psi_j=\psi^+_j(x,y,t)$ for $j=1,\ldots,M$ be a linearly independent set of solutions to the linear system 
\begin{equation}
i\pa_y\psi+\pa_x^2\psi=0,\quad \pa_t \psi+4\pa_x^3\psi=0,
\label{eq:linearsystem}
\end{equation}
and similarly let $\psi^-_j(x,y,t)$ be solutions to the conjugate system
$$
i\pa_y\psi-\pa_x^2\psi=0,\quad \pa_t \psi+4\pa_x^3\psi=0.
$$
Assume that all $\psi^{\pm}_j$ lie in $L^2((-\infty,x_0])$ with respect to the variable $x$ for any $x_0$, and let $c_{jk}$ be an arbitrary constant $M\times M$-matrix. Then the function
\begin{equation}
u(x,y,t)=2\pa_x^2\log \tau,\quad \tau(x,y,t)=\det\left[c_{jk}+\langle \psi^+_j,\psi^-_k\rangle\right],\quad
\langle \psi^+_j,\psi^-_k\rangle=\int_{-\infty}^x \psi_j^+(x',y,t)\psi_k^-(x',y,t)dx'
\label{eq:tau}
\end{equation}
is a solution of the KP-I equation~\eqref{eq:KPI}. To obtain real-valued solutions, we let $c_{jk}$ be real-valued, and we set $\psi^-_j=\opsi_j$.

It is customary to choose $c_{jk}=\delta_{jk}$ to ensure that the solution~\eqref{eq:tau} is non-singular; we call solutions of KP-I obtained in this way \emph{regular}. In this paper, however, we are more interested in the case $c_{jk}=0$; we call such solutions \emph{reduced}. We note that if the solutions $\psi_j$ are linearly independent, then the reduced $\tau$-function~\eqref{eq:tau} is the determinant of a Gram matrix, and hence the solution is nonsingular. We discuss the relationship between regular and reduced solutions of KP-I in Section~\ref{sec:regularvsreduced}, for now we note that the latter can be obtained from the former by setting $\psi^+_j=C\psi_j$, where $C$ is a real constant, and taking the limit $C\to +\infty$. It would also be interesting to consider solutions where the matrix $c_{jk}$ is nonzero but does not have maximal rank, however this is beyond the scope of our paper. 

In this paper, we restrict our attention to functions $\psi_j$ with finite spectral support. Fix a positive integer $N$, called the \emph{order} of the solution, and fix distinct eigenvalues $\la_1,\ldots,\la_N$ with positive real parts. Denote 
$$
\phi(x,y,t,\la)=\la x+i\la^2y-4\la^3t,
$$
and let $p_s(x,y,t,\la)$ denote the polynomial (homogeneous of degree $s$ in $x$, $y$, and $t$) defined by
$$
p_s(x,y,t,\la)=e^{-\phi(x,y,t,\la)}\pa_{\la}^se^{\phi(x,y,t,\la)},
$$
so that for example
$$
p_0=1,\quad p_1=x+2i\la y-12\la^2t, \quad
p_2=p_1^2+2iy-24\la t,\quad \ldots
$$
Any function of the form $\pa_{\la}^se^{\phi}=p_s e^{\phi}$ is a solution of~\eqref{eq:linearsystem}.

We now consider solutions of KP-I given by the tau-function~\eqref{eq:tau}, where the eigenfunctions $\psi_j$ are given by

\begin{equation}
\label{eq:psigeneral}    
\psi_j(x,y,t)=\sum_{n=1}^N\sum_{s=0}^S C_{jns}p_s(x,y,t,\la_{jn})e^{\phi(x,y,t,\la_{jn})}.
\end{equation}

The highest degree $S$ of a polynomial $p_s$ that occurs in any of the $\psi_j$ is called the \emph{depth} of the solution. The complex constants $C_{jns}$ are required to satisfy a non-degeneracy condition to ensure that the functions $\psi_j$ are linearly independent. We do not spell out this condition, and instead verify it in each particular example. 

An exhaustive classification of the solutions of KP-I obtained in this manner is far beyond the scope of this paper. Instead, our goal is to describe several interesting families of solutions that illuminate the behavior of the generic solution.

\begin{enumerate}

    \item \emph{Line-solitons.} The simplest solution of KP-I, called a \emph{line-soliton}, is the regular solution obtained from~\eqref{eq:tau} and~\eqref{eq:psigeneral} for $M=1$, $N=1$, and $S=0$, in other words by setting $\psi(x,y,t)=Ce^{\phi(x,y,t,\la)}$. This solution is a translation-invariant traveling wave, and a similar solution exists for KP-II. However, unlike the KP-II case, a line-soliton solution of KP-I is unstable (see~\cite{1993PelinovskyStepanyantsB,1975Zakharov}). 
    
    \item \emph{Rational solutions: lumps and multi-lumps.} A distinguishing feature of the KP-I equation is the existence of rational, spatially localized solutions, which are not known for KP-II. Consider the solution of KP-I given by~\eqref{eq:tau} and~\eqref{eq:psigeneral},  where each function $\psi_j$ is a polynomial multiple of a single exponential $e^{\phi(x,y,t,\la_j)}$ (the eigenvalues $\la_j$ corresponding to the $\psi_j$ may or may not be distinct). In this case the integral $\langle \psi^+_j,\psi^-_k\rangle$ occurring in~\eqref{eq:tau} is a polynomial multiple of $e^{(\la_j+\ola_k) x}$. In the regular case (when $c_{jk}=\delta_{jk}$), the $\tau$-function is a sum of distinct exponentials. However, in the reduced case (when $c_{jk}=0$), the $\tau$-function is a polynomial multiple of a single exponential term $\exp\sum(\la_j+\ola_j)x$, and the exponential disappears when taking the second logarithmic derivative. Therefore, the corresponding solution $u$ is a rational function of $x$, $y$, and $t$. These are the so-called \emph{lump} and \emph{multi-lump} solutions of KP-I. Corresponding to each distinct eigenvalue $\la_j$ there is a lump, or, more generally, a collection of lumps, whose number is related to the depth $S$. The lumps in each collection are either bounded or undergo anomalous scattering, while the collections of lumps corresponding to different $\la_j$ undergo normal scattering without phase shifts. Multilump solutions of KP-I were obtained in a number of papers (see for example~\cite{1993GorshkovPelinovskyStepanyants,2018HuHuangLuStepanyants,1978Krichever,1979Krichever,1994Pelinovsky,1998Pelinovsky,1993PelinovskyStepanyantsA,2017Stepanyants}). The most general Grammian form of the multilump solutions of KP-I was considered in~\cite{2018Chang}.  

    \item \emph{Lump chains.} In this paper, we are mostly concerned with reduced solutions of depth $S=0$, in other words when each function $\psi_j$ is a linear combination of exponentials. In order for the solution to be non-singular, we require $N\geq M$. As we will see, the corresponding reduced solution $u$ of KP-I is an arrangement of \emph{lump chains}, which are sequences of lumps moving along parallel trajectories (the group velocity of the chain is in general distinct from the velocity of the individual lumps). The time evolution of the underlying linear arrangement supporting the lumps is very similar to that of the line-soliton solutions of KP-II (see~\cite{2018Kodama}). However, the linear arrangements that can occur for lump chains are more general than those of KP-II line-solitons, and allow for various degenerate configurations such as parallel or superimposed lump chains. The regular solution of KP-I of depth $S=0$ and rank $M=1$ consists of a linear arrangement of lump chains interacting with a single line-soliton of KP-I. We give a detailed description of certain families of reduced lump chain solutions in Section~\ref{sec:lumpchains}, and we give a single example of a regular solution of depth $S=0$ in Section~\ref{sec:regularvsreduced}.
    
    \item \emph{Lumps and lump chains.} In Section~\ref{sec:regularvsreduced}, we also construct an example of a reduced solution of depth $S>0$ that is not rational. The solution consists of a chain of lumps that emits, at a certain moment of time, a single lump which propagates away from the chain. We conjecture that the general reduced solution of KP-I with depth $S>0$ consists of an arrangement of lump chains, with an additional number of individual lumps being either emitted or reabsorbed by the chains. The general regular solution consists of such an arrangement, and additional line-solitons. 
\end{enumerate}

\section{Lump chains: reduced solutions of depth $S=0$.}

\label{sec:lumpchains}

In this section, we consider solutions of KP-I given by~\eqref{eq:tau} that are reduced ($c_{jk}=0$), and where the auxiliary $\psi$-functions are given by~\eqref{eq:psigeneral} with depth $S=0$, in other words are sums of pure exponential terms with no polynomial multiples. We mostly focus on solutions of rank $M=1$, in other words having the form
\begin{equation}
u(x,y,t)=2\pa_x^2\log \tau,\quad\tau(x,y,t)=\int_{-\infty}^x|\psi(z,t,y)|^2\,dz,
\label{eq:rank1}
\end{equation}
where the $\psi$-function is a sum of $N$ exponentials. We will see that such a solution is an arrangement of linear \emph{lump chains}, with the individual lumps moving with constant velocity along the chains, and the entire assembly evolving with time. Such lump chain solutions of KP-I bear a strong resemblance to the well-known line-soliton solutions of KP-II, which are the subject of an elaborate combinatorial theory (see~\cite{2018Kodama}).

The $\psi$-function defining a lump chain solution of rank $M=1$ and order $N$ is defined by $2N$ complex parameters. It is convenient to introduce them as follows. Let 
$$
\la_n=a_n+ib_n,\quad \th_n=\rho_n+i\ph_n,\quad n=1,\ldots,N,
$$
be complex constants, where we assume that $0<a_1\leq a_2\leq \cdots\leq a_N$ and that $\la_n\neq \la_m$ for $n\neq m$. Define the functions
$$
\Phi_n(x,y,t)=\la_n x+i\la_n^2y-4\la_n^3t+\th_n,
$$
then the function
$$
\psi(x,y,t)=\sum_{n=1}^N \sqrt{2a_n}e^{\Phi_n(x,y,t)}
$$
satisfies the linear system~\eqref{eq:linearsystem}. Plugging $\psi$ into~\eqref{eq:rank1}, we obtain the following formula for the $\tau$-function:
\begin{equation}
\tau(x,y,t)=\sum_{n=1}^N e^{2F_n}+\sum_{n=1}^{N-1}\sum_{m=n+1}^N 2\mu_{nm}
e^{F_n+F_m}\cos (G_n-G_m-\ph_{nm}),
\label{eq:tauchainsoliton}
\end{equation}
where we have denoted
$$
F_n(x,y,t)=\Re \Phi_n(x,y,t),\quad G_n(x,y,t)=\Im \Phi_n(x,y,t),
$$
and the constants $\mu_{nm}$ and $\ph_{nm}$ are given by
$$
\mu_{nm}=2\sqrt{\frac{a_na_m}{(a_n+a_m)^2+(b_n-b_m)^2}},\quad \ph_{nm}=\tan^{-1}\left(\frac{b_n-b_m}{a_n+a_m}\right).
$$
The large-scale structure of the solution is governed by the linear functions
$$
F_{nm}(x,y,t)=F_n(x,y,t)-F_m(x,y,t)=A_{nm}x+B_{nm}y+C_{nm}t+D_{nm}=0,
$$
where
\begin{equation}
A_{nm}=\Re (\la_n-\la_m),\quad B_{nm}=-\Im(\la_n^2-\la_m^2),\quad C_{nm}=-4\Re(\la_n^3-\la_m^3),
\quad D_{nm}=\Re(\th_n-\th_m).
\label{eq:AB}
\end{equation}
We now describe the structure of the corresponding solution $u(x,y,t)$ for $N\geq 2$ (the solution is trivial for $N=1$). We note that adding a common complex constant to the $\theta_n$ multiplies $\tau$ by a real constant, and hence does not change $u$, therefore the solution is in fact determined by $2N-1$ complex parameters.

\subsection{Lump chain of rank $M=1$ and order $N=2$.} \label{subsec:order2} The solution of KP-I with $\tau$-function~\eqref{eq:tauchainsoliton} of rank $M=1$ and order $N=2$ is the basic building block for solutions of order $N\geq 3$, so we study it in detail. This solution is a linear traveling wave consisting of an infinite chain of lumps, and is analogous to the simple line-soliton solution of KP-II. In the $N=2$ case, the $\tau$-function~\eqref{eq:tauchainsoliton} can be simplified by factoring out the exponential term $e^{F_1+F_2}$. The corresponding solution of KP-I is given by
\begin{equation}
u(x,y,t)=2\frac{\pa^2}{\pa x^2}\log\left[\cosh(F_{21})+\mu_{12}\cos(G_2-G_1-\ph_{21})\right].
\label{eq:N=2u}
\end{equation}
The arguments of the $\cosh$ and $\cos$ functions are linear:
$$
F_{21}=A_{21}x+B_{21}y+C_{21}t+D_{21},\quad 
G_2-G_1-\ph_{21}=\al_{21} x+\be_{21} y + \ga_{21} t+\de_{21},
$$
where we have denoted
$$
\al_{21}=\Im(\la_2-\la_1),\quad \be_{21}=\Re (\la_2^2-\la_1^2),\quad
\ga_{21}=-4\Im(\la_2^3-\la_1^3),\quad \delta_{21}=\Im(\theta_2-\theta_1).
$$
Introduce the quantities 
$$
X=\frac{B_{21}\ga_{21}-C_{21}\be_{21}}{A_{21}\be_{21}-B_{21}\al_{21}},\quad Y=\frac{C_{21}\al_{21}-A_{21}\ga_{21}}{A_{21}\be_{21}-B_{21}\al_{21}},
$$
then the non-degeneracy condition $0<a_1\leq a_2$ implies that the vector $(X,Y)$ is nonzero, and that the denominators do not vanish. The functions $F_{21}=F_2-F_1$ and $G_2-G_1$, and therefore $u$, satisfy the differential equation $f_t+Xf_x+Yf_y=0$, hence $u(x,y,t)=u(x-Xt,y-Yt)$ is a traveling wave with the velocity vector $(X,Y)$. Furthermore, $u$ satisfies the stationary Boussinesq equation
$$
\left[-Xu_x-Yu_y+6uu_x+u_{xxx}\right]_x=3u_{yy}.
$$
For a fixed moment of time $t$, the solution~\eqref{eq:N=2u} is localized near the line $F_{21}=0$ in the $(x,y)$-plane. Indeed, away from the line, the argument of $\cosh$ has a large absolute value, so the $\tau$-function has a single dominant exponential term and hence $u$ is exponentially small. The normal direction vector $\bU_{21}=(A_{21},B_{21})$ of the line $F_{21}=0$ may be an arbitrary nonzero vector, and is not in general parallel to the velocity $(X,Y)$. In particular, if $a_1=a_2$ then the line $F_{21}=0$ is parallel to the $x$-axis, which cannot happen for a line-soliton of KP-II. The line $F_{21}=0$ propagates with the wave vector 
\begin{equation}
\bV_{21}=-\frac{C_{21}}{A_{21}^2+B_{21}^2} (A_{21},B_{21}),
\label{eq:groupvelocity}
\end{equation}
and is stationary if $C_{21}=0$ (which also cannot happen for a KP-II line-soliton). Note, however, that $C_{21}\neq 0$ if the line $F_{21}=0$ is vertical.

\begin{center}
\begin{figure}
\includegraphics[width=6.25in]{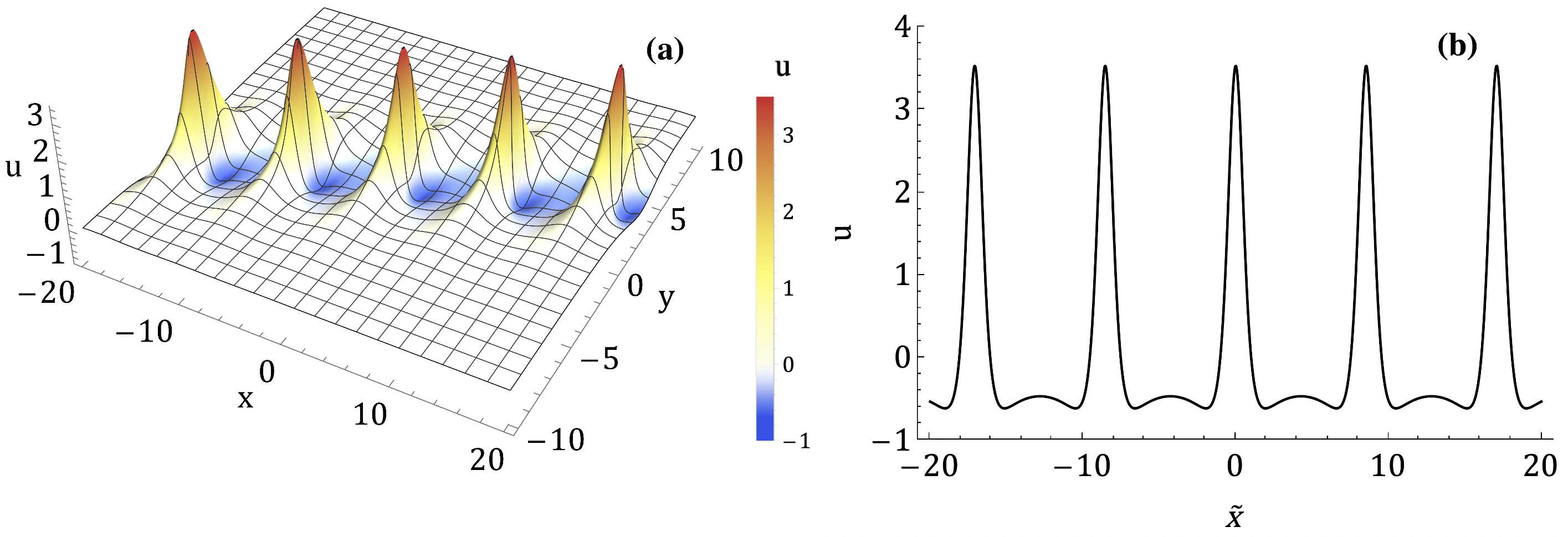}
\caption{Reduced lump chain of order $M=1$ and rank $N=2$, given by Equation~\eqref{eq:N=2u} with $\la_1=1/2+i/2$ and $\la_2=3/8-i/4$ at $t=0$.
 \textbf{(a)} $2$D profile of $u(x,y)$.
 \textbf{(b)} Amplitude of $u(\tilde{x},y)$ along the line $F_{12}=0$.}
\label{fig:lumpchain}
\end{figure}
\end{center}

Along the line $F_{21}=0$, the phase of the solution~\eqref{eq:N=2u} is determined by the argument of the cosine function. The solution is periodic along the line, and consists of a sequence of lumps (see Figure~\ref{fig:lumpchain}). The distance between two consecutive lumps is equal to
\begin{equation}
L_{21}=2\pi \frac{\sqrt{A_{21}^2+B_{21}^2}}{A_{21}\be_{21}-B_{21}\al_{21}}.
\label{eq:lumpspacing}
\end{equation}
The individual lumps propagate with velocity vector $(X,Y)$, which may be oriented arbitrarily relative the chain $F_{21}=0$. To see that the individual peaks are indeed KP-I lumps, we note that the distance $L_{21}$ between two consecutive lumps diverges as $\la_2\to \la_1$. Setting
$$
a_1=a-\ep,\quad b_1=b-\ep\mu,\quad
a_2=a+\ep,\quad b_2=b+\ep\mu, \quad\theta_1=\theta_2=0,
$$
in the limit $\ep\to 0$ we obtain (for $t=0$ and arbitrary $\mu$) the lump solution of KP-I:
\begin{equation}
u(x,y)=2\frac{\pa^2}{\pa x^2}\log [1+a^2(x-by)^2+a^4y^2]=\frac{4a^2(1-a^2(x-by)^2+a^4y^2)}{(1+a^2(x-by)^2+a^4y^2)^2}.
\label{eq:lump}
\end{equation}

The KP-I equation has infinitely many integrals of motion, the simplest being $\displaystyle\int_{-\infty}^{\infty} u(x,y,t)dx$ (in general, this integral is a linear function of $y$, but for our solutions it is in fact constant). It is easy to verify that for a lump chain of order $N=2$ we have 
$$
\frac{1}{4}\int_{-\infty}^{\infty} u(x,y,t)\,dx=A_{21}.
$$
We call the quantity $A_{21}$ the \emph{flux} of the lump chain.

We note that the integral $\displaystyle\int_{-\infty}^{\infty} u(x,y,t)\,dx$ is equal to zero for a one-lump solution~\eqref{eq:lump}, since $u(x,y)$ is the $x$-derivative of a rational function that vanishes at infinity. This agrees with the limiting procedure, since $A_{21}\to 0$ as $\la_2\to \la_1$. However, for the one-lump solution $u(x,y)$ given by~\eqref{eq:lump} the integral over the entire plane is nonzero:
$$
\int_{\RR^2}u(x,y)\,dx\wedge dy=\frac{4\pi}{a}>0.
$$
There is no contradiction here, since $u(x,y)$ does not vanish sufficiently rapidly as $x^2+y^2\to \infty$, and this improper integral cannot be evaluated using Fubini's theorem.

It has already been observed by a number of authors that a linear chain of lumps can occur as part of a solution of the KP-I equation. A chain of lumps appears in~\cite{1984ZhdanovTrubnikov}, and formula~\eqref{eq:N=2u} occurs in~\cite{1993Konopelchenko} (see p.~74), but is not analyzed in detail. In~\cite{1993PelinovskyStepanyantsB}, chains of lumps parallel to the $y$-axis are shown to result from the decay of an unstable line-soliton. Zaitsev~\cite{1983Zaitsev} developed a procedure for constructing stationary wave solutions of integrable systems out of spatially localized solitons, and constructed a lump chain for KP-I in this manner. Burtsev showed in~\cite{1985Burtsev} that a lump chain is unstable with respect to transverse perturbations, as is the case for a line-soliton. The development of the instability of the chain soliton was studied in~\cite{1993PelinovskyStepanyantsB}

\subsection{Lump chains of rank $M=1$ and order $N=3$.}

We now consider the reduced solutions $u(x,y,t)$ of KP-I with $\tau$-function given by~\eqref{eq:tauchainsoliton} in the case $N=3$. A generic solution of this form consists of three lump chains meeting at a triple point, and a number of degenerate configurations are also possible.

The $\tau$-function~\eqref{eq:tauchainsoliton} for $N=3$ consists of three purely exponential terms $e^{2F_1}$, $e^{2F_2}$, and $e^{2F_3}$, and three mixed terms. Introduce the normal vectors $\bU_{mn}=(A_{mn}, B_{mn})$, where $A_{mn}$ and $B_{mn}$ are given by~\eqref{eq:AB}. The normal vectors satisfy $\bU_{31}=\bU_{21}+\bU_{32}$, and their collinearity is controlled by the quantity
\begin{equation}
\eta_{123}=A_{21}B_{31}-A_{31}B_{21}=a_1b_1(a_2-a_3)+a_2b_2(a_3-a_1)+a_3b_3(a_1-a_2).
\label{eq:eta}
\end{equation}
For generic values of $\la_1$, $\la_2$, and $\la_3$ we have $\eta_{123}\neq 0$, and no two of the three vectors $\bU_{mn}$ are collinear. In this case, the $(x,y)$-plane is partitioned, for fixed $t$, into three sectors meeting at a triple point. In each sector, one of the pure exponential terms $e^{2F_m}$ is dominant, and the solution $u$ is exponentially small. Along the boundary of two sectors, given by the equation $F_{mn}=0$, two of the exponentials $e^{2F_m}$ and $e^{2F_n}$ are equal and are comparable to the mixed exponential term containing $e^{F_m+F_n}$. The triple point is given by the equation $F_1=F_2=F_3=0$, moves linearly with $t$, and passes through $(0,0)$ at $t=0$ if the phases $\th_m$ are purely imaginary. 

\begin{figure}[!t]
\includegraphics[width=6.5in]{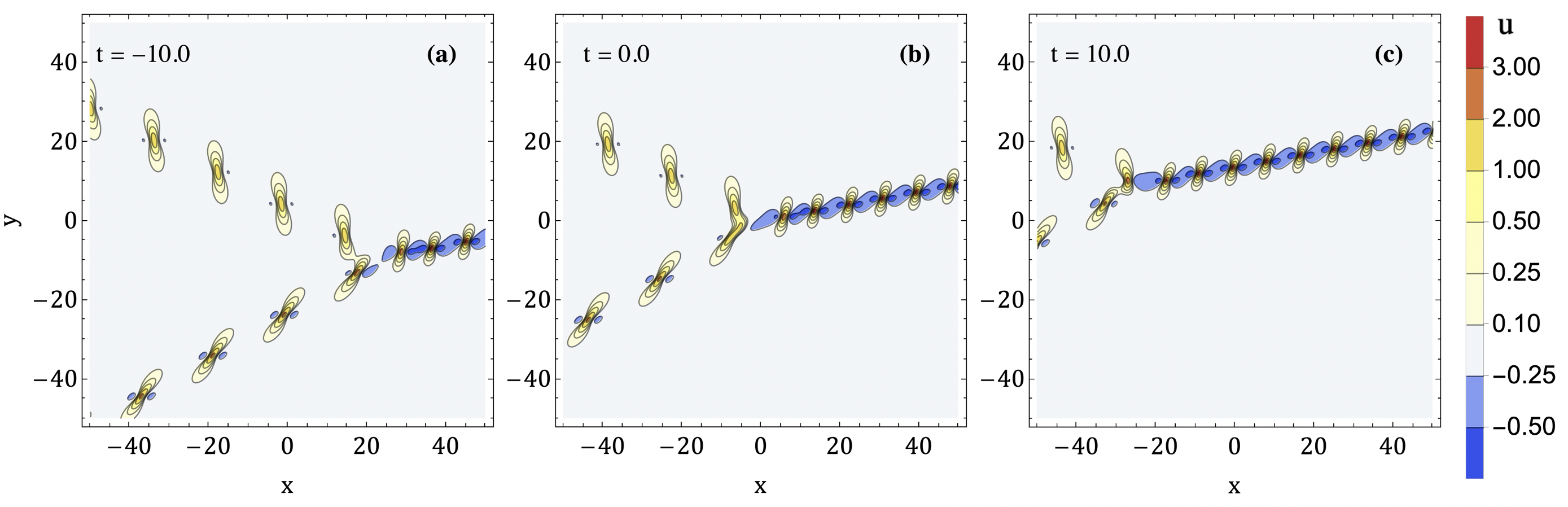}
\caption{ $N=3$ lump chain with $\la_1=1/2+i/2$, $\la_2=3/8-i/4$  and $\la_3=1/4+i/8$ at different moments of time.}
\label{fig:triplepoint1}
\end{figure}

\begin{figure}
\includegraphics[width=6.5in]{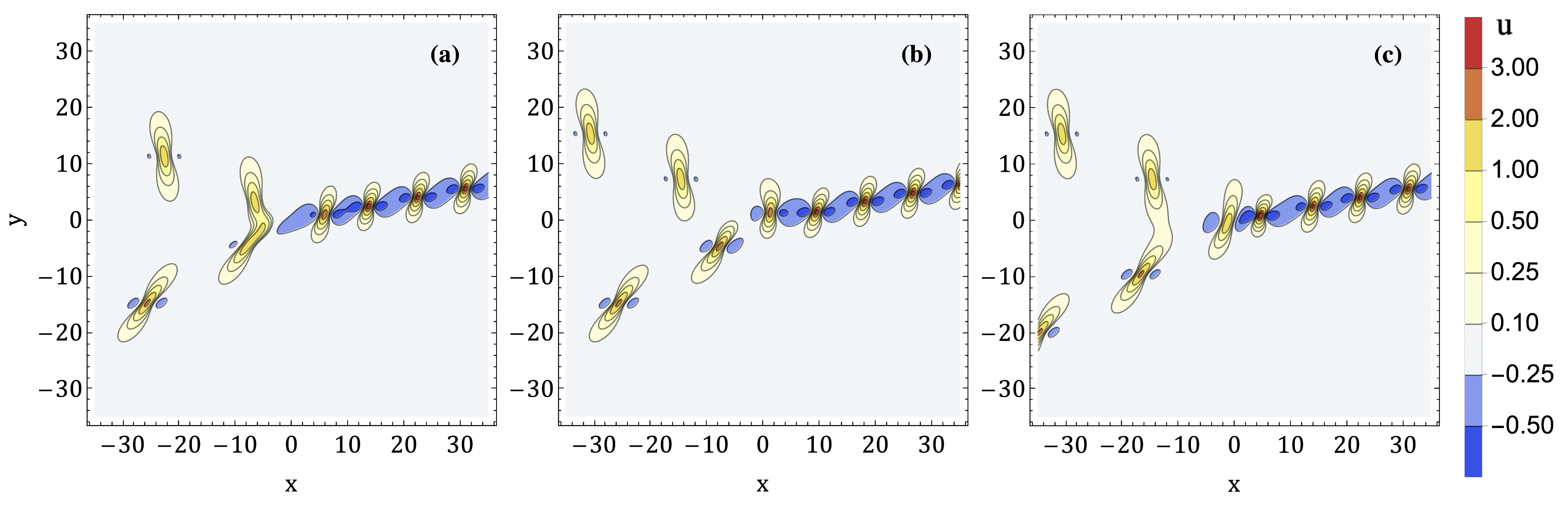}
\caption{ $N=3$ lump chain with $\la_1=1/2+i/2$, $\la_2=3/8-i/4$  and $\la_3=1/4+i/8$ at $t=0$, and with different relative phases. 
\textbf{(a)} $\delta_{12} = 0$ and $\delta_{13}= 0$ \textbf{(b)} $\delta_{12} = \pi$ and $\delta_{13} = 0$ \textbf{(c)} $\delta_{12}= 0$ and $\delta_{13} = \pi$.}
\label{fig:triplepoint2}
\end{figure}

The solution itself is localized on the boundaries of the sectors. Along the boundary $F_{mn}=0$, the solution can be approximated by an order $N=2$ solution described in Subsection~\ref{subsec:order2}. In other words, it is a lump chain with normal vector $\bU_{mn}$, which we call an $[m,n]$-chain. Depending on the values of the spectral parameters, there are two possibilities. In the first, shown on Figure~\ref{fig:triplepoint1}, the $[2,1]$- and $[3,2]$-lump chains meet at the triple point. The individual lumps from the two chains interlace one by one and form the new $[3,1]$-chain.  Conversely, the lumps on the $[3,1]$-chain may split at the triple point into two new chains. In either case, individual lumps are preserved, and the fluxes of the three chains satisfy the local conservation law $A_{31}=A_{21}+A_{32}$. In addition to the orientation of the chains, the position of the triple point, and the velocities of the lumps along the chains, there are two free parameters that determine the solution, namely the relative phases $\delta_{mn}=\Im(\theta_m-\theta_n)$. Figure~\ref{fig:triplepoint2} shows the solution for three different sets of values of the relative chain phases.

\begin{figure}[!h]
\includegraphics[width=6.5in]{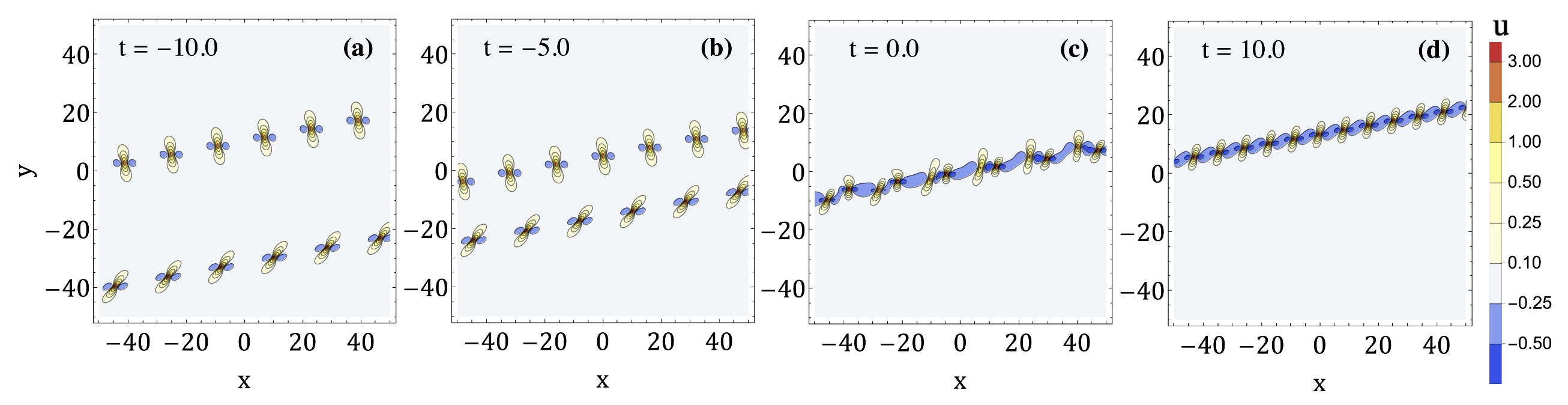}
\caption{Two parallel lump chains merging into one: $N$=3, $\la_1=1/2+i/2$, $\la_2=3/8-i/4$ and $\la_3=3/7+i/8$ at different moments of time. \textbf{(a)} $t=-15.0$ \textbf{(b)} $t=-10.0$ \textbf{(c)} $t=-5.0$ \textbf{(d)} $t=0.0$ \textbf{(e)} $t=10.0$ \textbf{(a)} $t=20.0$ }
\label{fig:parallelchainsmerging}
\end{figure}

There are additionally a number of degenerate configurations, corresponding to $\eta_{123}=0$. In this case the vectors $\bU_{21}$, $\bU_{31}$, and $\bU_{32}$ are collinear, and hence so are the lines $F_{21}=0$, $F_{31}=0$ and $F_{32}=0$. Depending on the values of the $\la_m$, for fixed $t$, either the $(x,y)$-plane consists of half-planes in which $e^{2F_1}$ and $e^{2F_3}$ are dominant, or there is an additional intermediate strip in which $e^{2F_2}$ is dominant. For generic $\la_1$, $\la_2$, and $\la_3$ (satisfying $\eta_{123}=0$), the solution may consist of two parallel $[2,1]$- and $[3,2]$-chains merging into the $[3,1]$-chain, as shown on Figure~\ref{fig:parallelchainsmerging}. The opposite case is also possible: a single $[3,1]$-chain may, at a certain moment of time, split into two lump chains, both parallel to the original chain. A similar splitting process was observed in~\cite{1993PelinovskyStepanyantsB}.

Imposing the additional condition $A_{21}C_{31}-A_{31}C_{21}=0$, we obtain a further degeneration: the three lines $F_{21}=0$, $F_{31}=0$, and $F_{32}=0$ that can support the chains move are not only parallel, but move with equal velocity. In this case, the solution consists either of two parallel lump chains propagating at a fixed distance, or of a single lump chain (in the latter case, the solution may be visually indistinguishable from a lump chain of order $N=2$). 

\begin{figure}[!h]
\includegraphics[width=6.5in]{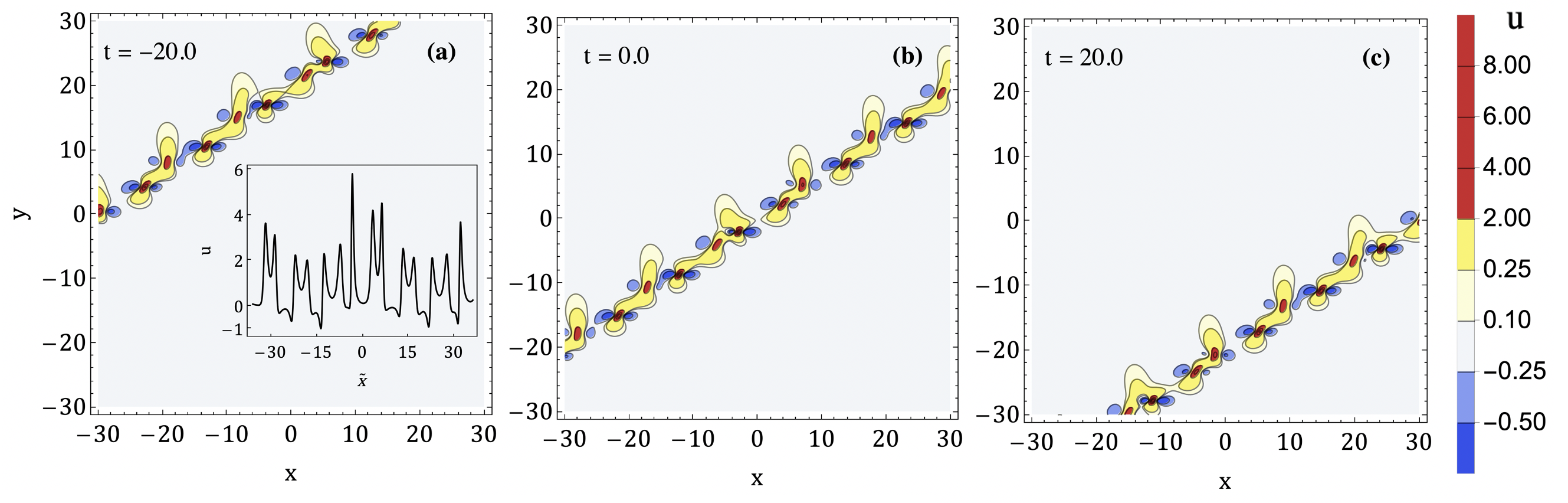}
\caption{\small Quasi-periodic lump chain of order $M=1$ and rank $N=3$, $\la_1=1+i/2$, $\la_2=1/4-i/4$ and $\la_3=0.443+0.186i$, at times $t=-20.0$,  $t=0.0$, and $t=20.0$. Inset in \textbf{(a)} shows amplitude along the quasi-periodic lump chain.
}
\label{fig:doublechain}
\end{figure}

Finally, it is possible that the three lines $F_{21}=0$, $F_{31}=0$, and $F_{32}=0$ are the same for all values of $t$. The three lump chains merge into a complex, periodic or quasi-periodic chain supported along the common line, which propagates linearly (see Figure~\ref{fig:doublechain}). 

\subsection{Lump chains of rank $M=1$ and order $N\geq 4$.}

We now discuss the general form of the solution~\eqref{eq:tauchainsoliton} for arbitrary $N$, which is determined by the spectral parameters $\la_n$. The $\tau$-function is a sum of purely exponential terms $e^{2F_n}$ and mixed exponentials $e^{F_n+F_m}$ with trigonometric multipliers. For fixed $t$, the shape of the solution is determined by the relative values of the $F_n$: if one exponential $e^{2F_n}$ is dominant in the $\tau$-function, then the solution $u$ is exponentially small (a mixed term containing $e^{F_n+F_m}$ cannot be the only dominant term). Hence the $(x,y)$-plane is divided into finitely many polygonal regions, in the interior of which a single term $F_n$ is dominant. In fact, a given term $F_n$ may in general fail to be dominant anywhere, so there may be fewer than $N$ regions, and there may be only two. On the boundary of $F_{nm}=F_n-F_m=0$ of two regions, where $F_n=F_m$ and all other $F_k$ are significantly smaller, the solution can be approximated by an order $N=2$ solution described in Subsection~\ref{subsec:order2}. In other words, it is a lump chain with normal vector $\bU_{nm}=(A_{nm},B_{nm})$, which we call an $[n,m]$-chain. Hence the $(x,y)$-plane decomposes into finitely many polygonal regions with lump chains along the boundaries, and the entire arrangement evolves linearly with $t$. The structure of the lump chains closely resembles the arrangement of line-solitons in KP-II (see~\cite{1978AnkerFreeman,2018Kodama}).

\begin{figure}[!h]
\includegraphics[width=6.5in]{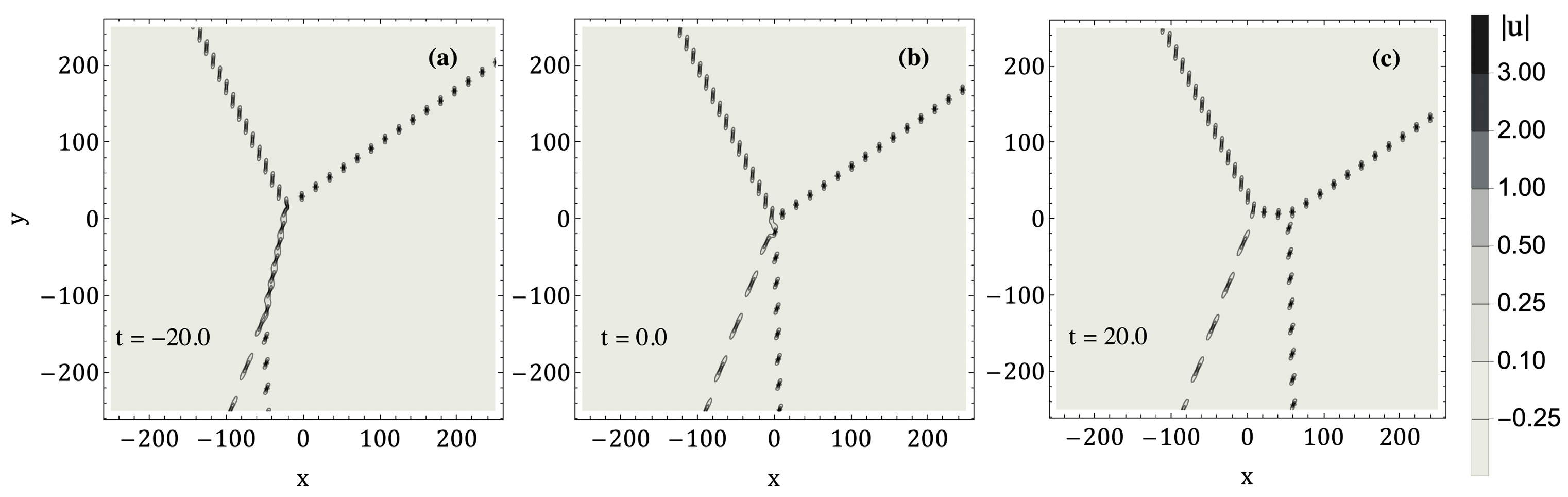}
\caption{H-shaped arrangement of chains of order $M=1$ and rank $N=4$, with eigenvalues $\la_1=17/32+i/8$, $\la_2=3/8-i/8$, $\la_3=11/32+i/5$, $\la_4=3/32+i/4$ and times $t=-20.0$, $t=0$ and $t=20.0$.}
\label{fig:N=4exampleH}
\end{figure}

We do not develop a general theory describing the line structure of the solutions. Instead, we give two generic examples of order $N=4$, and discuss the possible degenerate behavior. The first example, shown on Figure~\ref{fig:N=4exampleH}, may be called an $H$-configuration. It consists two triple points that are separated by a lump chain bridge. The bridge contracts and disappears at $t=0$, and the triple points scatter along a different bridge. A similar configuration appears in the KP-II equation (see~\cite{1978AnkerFreeman,2018Kodama}).

\begin{figure}[!h]
\includegraphics[width=6.5in]{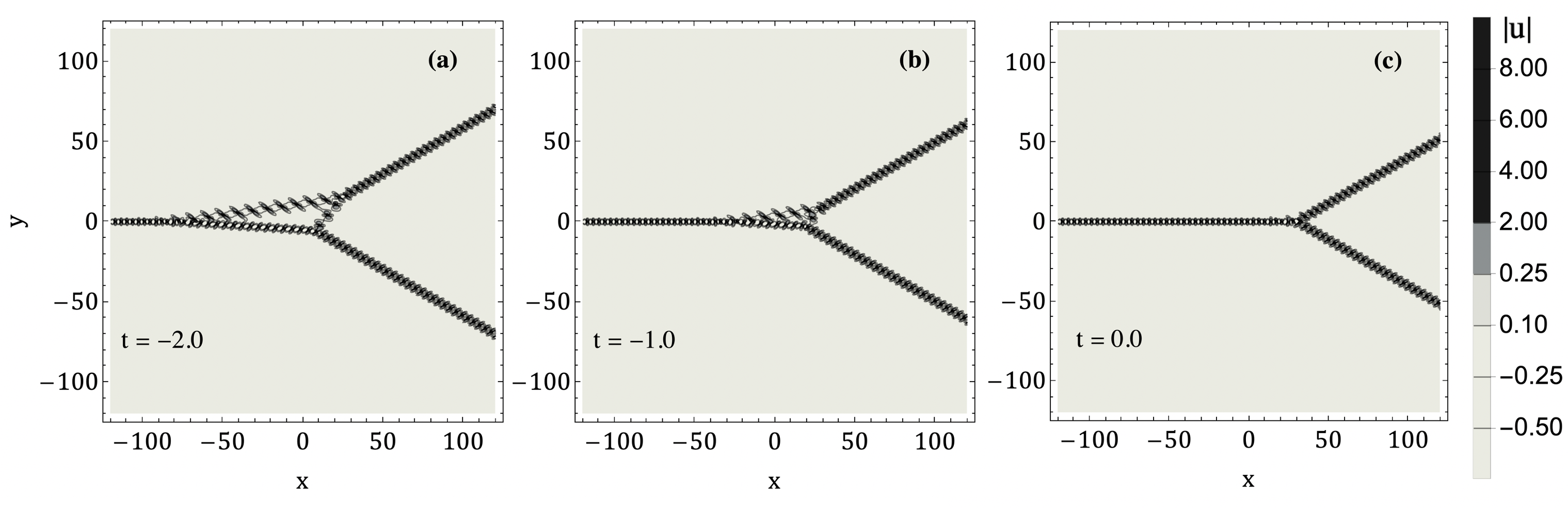}
\caption{Triangular arrangement of chains of order $M=1$ and rank $N=4$, with eigenvalues $\la_1=1$, $\la_2=1/2+i\sqrt{3}/2$, $\la_3=1/2-i\sqrt{3}/2$, $\la_4=1/\sqrt{3}-i/(2\sqrt{3})$, and times $t=-2.0$, $t=-1.0$ and $t=0.0$.
}
\label{fig:N=4exampletriangle}
\end{figure}

The second example, shown on Figure~\ref{fig:N=4exampletriangle}, has three triple points bounding a finite triangular region. The region shrinks and disappears at $t=0$, and the solution henceforth resembles a solution of order $N=3$. This configuration can be reversed in time, with a triangular region appearing out of a triple point. We stress that both these examples are generic, in other words the structure of the lump chains does not change under small perturbations of the $\la_n$.

The reader may recognize that the structure of lump chain solutions of KP-I is very similar to the structure of line-soliton solutions of KP-II. We point out that the line structure in the KP-I case may in fact be more complex. Specifically, the following kinds of behavior, all of them forbidden for KP-II line-solitons, can occur for KP-I lump chains of rank $M=1$ and order $N$.

\subsubsection{Generic solutions: the number of chains at infinity and forbidden configurations.} We first consider the case when the eigenvalues $\la_n$ are sufficiently generic. A natural first question is to determine the linear configurations of chains that may occur, in particular, the number of lump chains extending to infinity. An order $N$ line-soliton of KP-II always has $N$ solitons extending to infinity, but a generic order $N$ solution of KP-I may have anywhere between 3 and $N$ infinite lump chains. Similarly, certain configurations of lines are forbidden for KP-II line-solitons but may occur for KP-I lump chain solutions. For example, the solution given on Figure~\ref{fig:N=4exampletriangle} has $N=4$ and three infinite chains, and represents a line arrangement that cannot occur in KP-II (see Exercise 4.6 in~\cite{2018Kodama}).

\subsubsection{Degenerate solutions: stable points, parallel chains, and higher order chains.} Various degenerate configurations may be achieved by imposing appropriate conditions on the eigenvalues $\la_n$. The triple points where lump chains meet may be stationary relative to one another, and may even coincide for all times, producing stable quadruple points and points of higher multiplicity. A solution may have sets of parallel lump chains, in which case the number of chains at infinity may be greater than the order $N$. Finally, lump chains may coincide, producing quasiperiodic chains of higher order.

\subsection{Lump chains of rank $M\geq 2$.}

\begin{figure}[!h]
\includegraphics[width=6.5in]{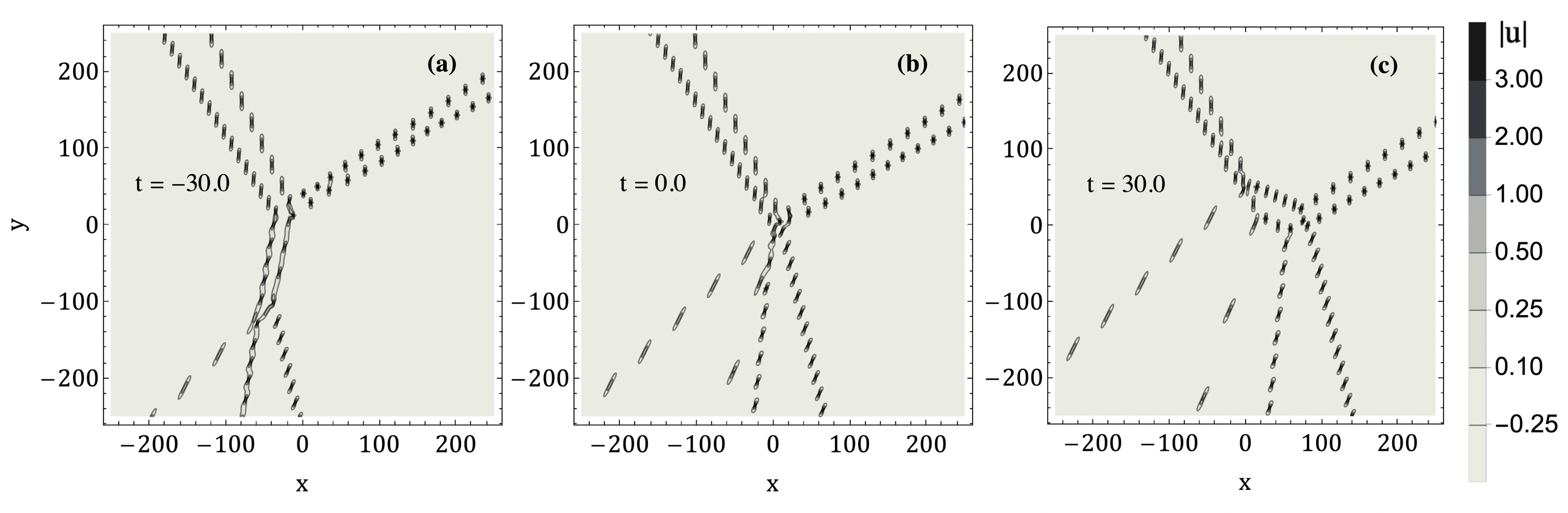}
\caption{The time evolution of a rank 2 order 4 solution with eigenvalues $\la_{11}=1/2+i/8$, $\la_{12}=3/8-i/8$, $\la_{13}=1/4+i/5$, $\la_{14}=1/8+i/4$, $\la_{21}=4/9+i/9$, $\la_{22}=3/9-i/9$, $\la_{23}=2/9+i/3$, $\la_{24}=1/9+i/7$}
\label{fig:rank2}
\end{figure}

The structure of reduced solutions of KP-I of depth $S=0$ and higher rank $M\geq 2$ is broadly similar to the $M=1$ case. The $\tau$-function~\eqref{eq:tau} is a sum of purely exponential terms and mixed terms involving trigonometric multipliers. For a given moment of time $t$, the $(x,y)$-plane is partitioned into finitely many polygons, and this decomposition evolves linearly with time. Polygonal regions may appear and disappear at certain moments of time. The boundaries of the polygons support lump chains, and the total flux of the lump chains arriving at a given vertex is equal to the flux of the chains that are leaving. In degenerate cases, there may be coinciding polygonal boundaries supporting quasiperiodic superpositions of lump chains. We give a single example of such a solution with rank $M=2$ and order $N=4$ in Figure~\ref{fig:rank2}.

\section{Regular solutions and solutions of depth $S>0$: line-solitons and individual lumps.}
\label{sec:regularvsreduced}

\begin{figure}[!h]
\includegraphics[width=6.5in]{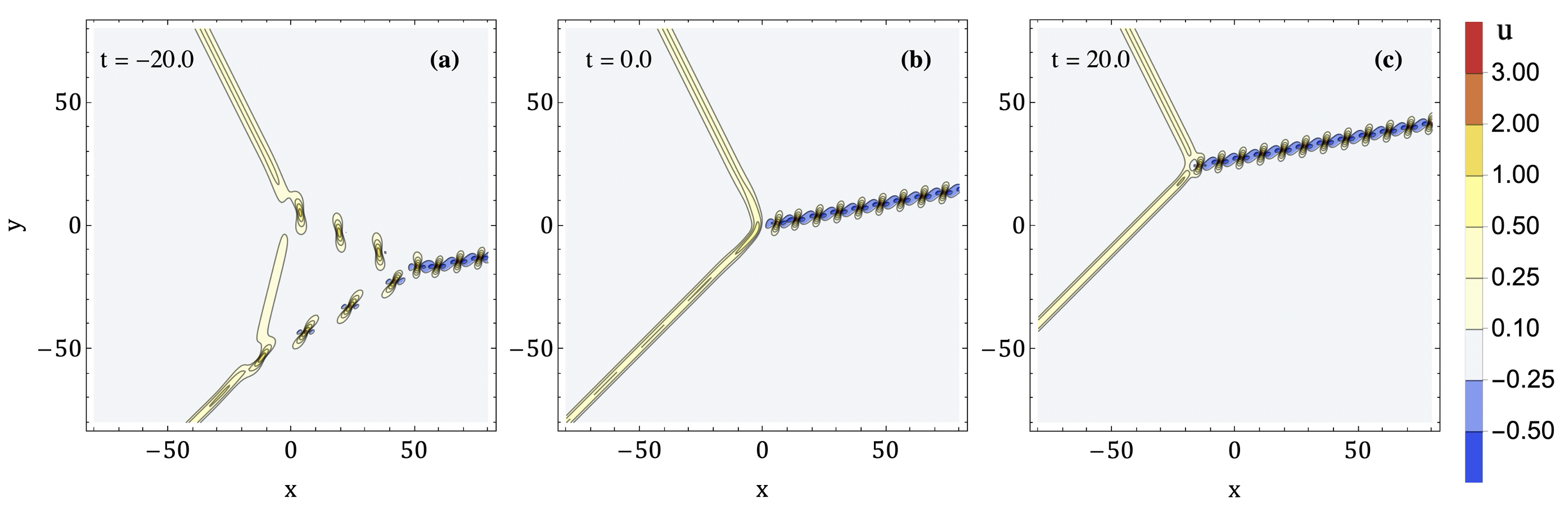}
\caption{Regular solution of rank $M=1$ and order $N=3$, with eigenvalues $\la_1=1/2+i/2$, $\la_2=3/8-i/4$, and $\la_3=1/4+i/8$, at different moments of time.}
\label{fig:laum_plus_chain}
\end{figure}

We now discuss the relationship between regular and reduced solutions of depth $S=0$, and solutions of positive depth. We first consider regular solutions, and for simplicity restrict our attention to rank $N=1$. The $\tau$-function of such a solution is nearly identical to that of the reduced solution~\eqref{eq:tauchainsoliton}, and has the form
\begin{equation}
\tau(x,y,t)=1+\sum_{n=1}^N e^{2F_n}+\sum_{n=1}^{N-1}\sum_{m=n+1}^N 2\mu_{nm}
e^{F_n+F_m}\cos (G_n-G_m-\ph_{nm}).
\label{eq:tauregular}
\end{equation}
As before, the $(x,y)$-plane is partitioned into polygonal regions, in each of which one of the terms in~\eqref{eq:tauregular} is dominant. However, there is now a new region, on which the dominant term in the $\tau$-function is the constant 1. Since $a_n=\Re \la_n>0$, this region contains, for a given fixed $y$, all points $(x,y)$ with sufficiently large negative $x$. Inside this region the $\tau$-function is approximately constant, and the solution $u$ is exponentially small. At the boundary of this region, the two dominant terms in the $\tau$-function are the 1 and one of the exponentials $e^{2F_n}$. Hence the boundary of the region where 1 dominates is a line-soliton of KP-I, instead of a lump chain. In other words, the solution consists of an infinite line-soliton of KP-I coupled with an arrangement of lump chains (see Figure~\ref{fig:laum_plus_chain}).

It is possible to degenerate a regular solution to a reduced solution by replacing the $1$ in Equation~\eqref{eq:tauregular} with an $\varepsilon$ and taking the limit $\varepsilon\to 0$. The line-soliton occurs on the boundary of the region where the $\varepsilon$ is the dominant term, and this region moves in the negative $x$-direction as $\varepsilon \to 0$. In the limit, the line-soliton disappears to infinity, and we are left with a solution consisting entirely of lump chains. Therefore, the limiting procedure that produces reduced solutions out of regular solutions has the effect of removing the line-soliton and isolating the lump chain structure. 

\begin{figure}[!h]
\includegraphics[width=6.5in]{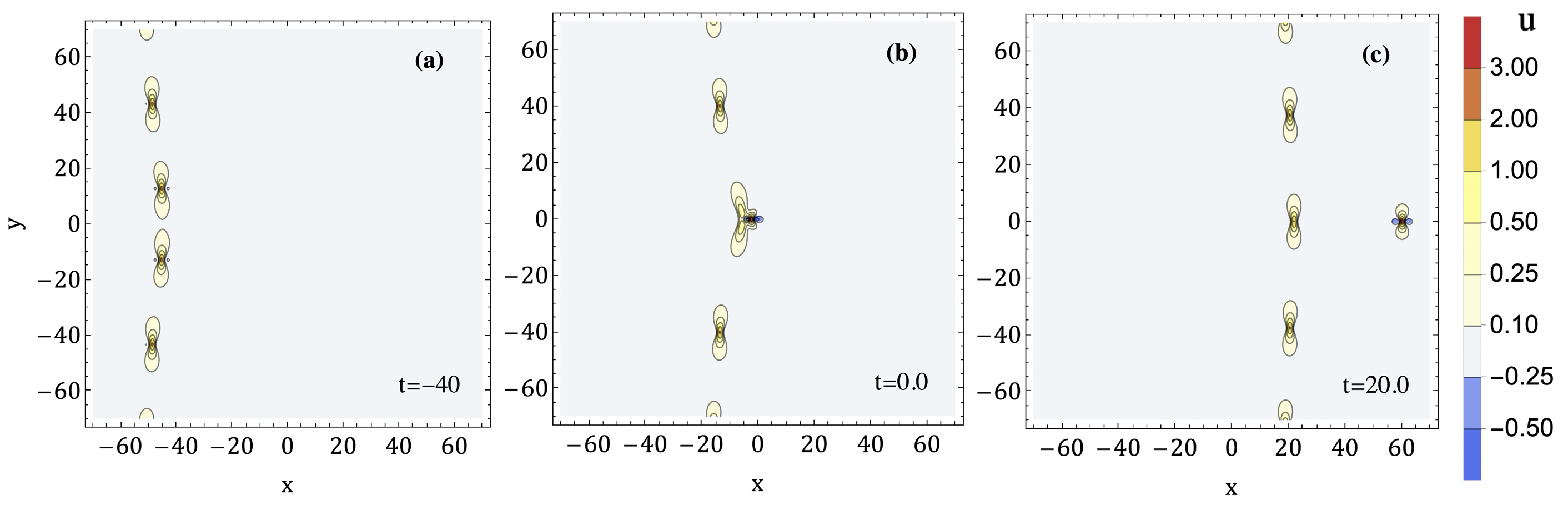}
\caption{A lump chain radiates an individual lump, which propagates away. $\la_1=1/4$, $\la_2=1/2$ with multiplicities of 1 and 2, respectively. }
\label{fig:lumpradiation}
\end{figure}

We also briefly consider the structure of reduced solutions of depth $S>0$. Consider again the general form of the $\tau$-function~\eqref{eq:tau}, where $c_{jk}=0$ and the $\psi_j$ are given by~\eqref{eq:psigeneral}. As discussed in Section~\ref{sec:Grammian}, the $\tau$-function is rational if each $\psi_j$ is a polynomial multiple of a single exponential term $e^{\phi(x,y,t,\la_j)}$. The corresponding solution is localized in the $(x,y)$-plane and represents the normal (if all $\la_j$ are distinct) or anomalous scattering of lumps, or even bound states of lumps. We now consider what happens in general, when each $\psi_j$ is a multiple of several exponentials. For sufficiently large $x$ and $y$, the polynomial terms are negligible compared to the exponentials, and the $\psi_j$ can be assumed to be purely exponential. Hence the solution can be assumed to have depth $S=0$ and is an arrangement of lump chains, as described in Section~\ref{sec:lumpchains}. In the finite part of the $(x,y)$-plane, however, the polynomial terms in the $\psi_j$ produce individual lumps. Hence, the overall structure of the solution is an arrangement of lump chains interacting with finitely many individual lumps: a lump chain may emit or absorb an individual lump, and the lumps may scatter on one another. A detailed classification of such solutions appears to be a challenging combinatorial problem. In Figure~\ref{fig:lumpradiation}, we give a single example of such a solution, consisting of a lump chain emitting an individual lump. We note that the local number of lumps is conserved: two lumps from the chain meet and scatter, with one lump propagating away and the other filling the resulting gap in the chain. Figure~\ref{fig:lumpradiation} gives an example of such a solution with rank $M=1$, order $N=2$, and depth $S=1$, with the $\psi$-function given by
$$\psi(x,y,t)=e^{\frac{1}{4} (-2 t+2 x+i y)} (-3 t+x+i y+1)+e^{\frac{1}{16} (-t+4 x+i y)}.
$$


\section{Summary and conclusion}

We have constructed a new family of lump chain solutions of the KP-I equations using the Grammian form of the $\tau$-function. A simple lump chain consists of an infinite line of equally spaced lumps. The lumps propagate with equal velocity, which is in general distinct from the group velocity of the line. The general solution consists of an evolving polyhedral arrangement of lump chains. At a point where three or more lump chains meet, the individual lumps from the incoming chains are redistributed along the outgoing chains, with the number of lumps being locally conserved. The linear structure of the solutions is very similar to that of the line-soliton solutions of KP-II. However, various degenerate configurations may occur for KP-I lump chains that cannot occur for KP-II line-solitons: parallel chains, chains of equal velocity, quasiperiodic superimposed chains, stable points of high multiplicity, and forbidden polyhedral configurations. We have also constructed more general solutions of KP-I using the Grammian method. Such solutions consist of an arrangement of lump chains as described above, together with line-solitons and individual lumps that are emitted and/or absorbed by the lump chains. A detailed classification of the solutions of KP-I that may be obtained by the Grammian method is an interesting and difficult problem, and is beyond the scope of this paper. We plan to return to this problem in future work.

\section{Acknowledgments}

The work of the second and fourth author on the third chapter was supported by the Russian Science Foundation (Grant No. 19-72-30028).

\end{document}